\documentstyle[12pt]{article}
\newcommand{\newsection}{
\setcounter{equation}{0}
\section}
\newcommand{\rf}[1]{(\ref{#1})}
\newcommand{\beq}{\begin{equation}}
\newcommand{\eeq}{\end{equation}}
\newcommand{\bea}{\begin{eqnarray}}
\newcommand{\eea}{\end{eqnarray}}

\renewcommand{\l}{\lambda}

\renewcommand{\a}{\alpha}
\newcommand{\n}{\nu}
\newcommand{\m}{\mu}

\newcommand{\oh}{\frac{1}{2}}
\newcommand{\oq}{\frac{1}{4}}

\newcommand{\dg}{\dagger}

\begin{document}
\topmargin 0pt
\oddsidemargin 5mm
\headheight 0pt
\headsep 0pt
\topskip 9mm

\addtolength{\baselineskip}{0.20\baselineskip}
\hfill    NBI-HE-92-52

\hfill July 1992
\begin{center}

\vspace{36pt}
{\large \bf
Topography of the hot sphaleron Transitions}
\end{center}

\vspace{36pt}

\begin{center}
{\sl J. Ambj\o rn } and
{\sl K. Farakos}\footnote{Supported by an EEC Fellowship.\\
Permanent address: National Technical University of Athens, Depart. of Physics,
GR 157 73, Athens, Greece.}

\vspace{12pt}

 The Niels Bohr Institute\\
Blegdamsvej 17, DK-2100 Copenhagen \O , Denmark\\

\vspace{24pt}

\end{center}

\vfill

\begin{center}
{\bf Abstract}
\end{center}

\vspace{12pt}

\noindent
By numerical simulations in {\it real time}
we provide evidence in favour of sphaleron like
transitions in the hot, symmetric phase of the electroweak theory.
Earlier performed observations of a change in the Chern-Simons number
are supplemented with a measurement of the lowest eigenvalues of the
three-dimensional staggered fermion Dirac operator and observations
of the spatial extension of energy lumps associated with the transition.
The observations corroborate on the interpretation of the change in
Chern-Simons numbers as representing  continuum physics, not lattice
artifacts. By combining the various observations it is possible to follow
in considerable detail the time-history of thermal fluctuations of
the classical gauge-field configurations responsible for the
change in the Chern-Simons number.

\vspace{24pt}

\vfill

\newpage

\newsection{Introduction}

It is by now well known that the fermion number is not conserved in the
standard electroweak theory due to the anomaly of the fermionic current
and the periodic vacuum structure in the non-abelian gauge theories
\cite{hooft}. Although the amplitude for this process is exponentially small
at zero temperature a great amplification can occur at high temperature
\cite{krs}. The reason is that the fermion-number violating processes
at zero temperature can occur only by tunneling from one classical
vacuum to another, the two being connected by a large gauge transformation
with winding number different from zero. At higher temperature it is possible
to move  between the different gauge vacua by classical thermal
fluctuations. The energy barrier which separates the vacua has a height
determined by the sphaleron energy $E_{sph}$, which is the energy of the
static classical solution to the electroweak theory corresponding to the
lowest lying saddle point between two neighbour vacua \cite{manton}.
If the temperature $T$ is so ``low'' that the symmetry is broken
we have
\beq
E_{sph} = \frac{2M_w(T)}{\a_w} B(\l/\a_w)   \label{1.1}
\eeq
where $B(\l/a_w)$ is a factor of order one, which varies only slowly
when the ratio between the Higgs coupling $\l$ and $\a_w=g_w^2/4\pi$
changes from zero to infinity. In \rf{1.1} we have included a possible
temperature dependence in the $W$-mass $M_w$. As long as
$x\equiv E_{sph}/T << 1$ we can trust a one loop calculation and we
get that the transition probability per  volume and time
for moving from one vacuum to a neighbour one is given by \cite{al,bs}
\beq
\Gamma = 0.007 (\a_w T)^4 x^7 e^{-x}.  \label{1.2}
\eeq

This calculation is strictly speaking only valid as long as
the Boltzmann factor $e^{-x}$ dominates the prefactor $x^7$ from
zero modes of the sphaleron. It is natural to expect that
\rf{1.2} extrapolates to
\beq
\Gamma = \kappa (\a_w T)^4 \label{1.3}
\eeq
when the symmetry is restored ($M_w(T) =0$) and $x$ is formally zero
\cite{al,ks,mls}, but obviously we can get no information about the
non-perturbative constant $\kappa$ from \rf{1.2} and analytical methods have
until now failed in the symmetric phase due to infrared divergences in the
high temperature expansion. The constant $\kappa$ was
determined by real time computer simulations of the $SU(2)$
gauge-Higgs system and the result was that $\kappa \sim O(1)$ \cite{aaps}.

The cosmological implication of $\kappa \sim O(1)$ is that any  $B+L$,
the baryon plus lepton number, generated at $GUT$ temperature, will be
washed out before we reach the electroweak transition. In theories
where $B-L$ are strictly conserved the problem of explaining the
baryon asymmetry observed in the universe is then pushed to
the electroweak transition temperature.

Since the implications of $\kappa \sim O(1)$ are important
 we have felt a need to corroborate on the claim in \cite{aaps} that the
configurations observed by computer simulations are really to be
identified with continuum like configurations which change the
Chern-Simon numbers by units of one. The only thing done in
\cite{aaps} was to measure the ``naive'' lattice definition of
$\int_0^t \int d^3x F\tilde{F}$  as a function of time when it
developed according to the classical equations of motion, starting
with a hot configuration above the electroweak transition temperature
$T_c$. Whenever a change in the Chern-Simons number compatible
with unity and followed by some kind of plateau was observed
it was classified as a ``sphaleron'' like transition\footnote{Of course there
are no genuine sphalerons for $T > T_c$. We use the expression
``sphaleron-like configuration'' for any configuration where the
gauge field has the same qualitative features as the sphaleron:
The configuration is the one in a sequence of
configurations where one of the eigenvalues of the Dirac operator is zero
and it is one where there is a clear lump of energy located in a considerable
region of space.}. The rather rapid change of Chern-Simons number observed
in \cite{aaps} could potentially be due to lattice dislocations rather
than genuine extended field configurations with a ``topological''
interpretation.
We have in this paper tried to verify the ``continuum nature'' of the lattice
configurations in two ways:
By measuring the eigenvalues $E_n(t)$  of the three dimensional Dirac operator
at time $t$  and observing that one of them dives to zero
within the time period where the
Chern-Simon number changes rapidly by one unit, and further by measuring the
energy density as a function of time for the gauge field configurations.
The approach is from this point of view similar to the work analyzing the
monopole-like configurations in lattice QCD \cite{ls,hands}, since these
 should also be thought of as three dimensional  configurations.

\newsection{The model and the simulation}

The main approximation in the present work is the use of purely classical
thermodynamics. It is based on the observation that the energy of the
sphaleron is of order $M_w/\a_w$ while its extension is of order
$M_w^{-1}$. For $T >T_c$ we expect similar characteristica of the
sphaleron-like configurations: their energy will be of order $T$ but
their extension of order $(\a_wT)^{-1}$. Due to the smallness of $\a_w$ the
characteristic momenta of thermal fluctuations which form such sphaleron-like
configurations are therefore much smaller than the generic quantum
fluctuations of the hot plasma which are of order $T$. Hence the
sphaleron-like fluctuations decouple from the quantum fluctuations and
one might expect that the thermal fluctuations responsible for the change
in Chern-Simons number are well described by classical physics for
temperatures  above $T_c$.

The idea is therefore to start out in a classical configuration\footnote{
Throughout this paper we assume that we can ignore the hypercharge sector,
which means that we effectively work with the Weinberg angle $\theta_w=0$.}
\beq
\{A^a_i,\phi^a;E^a_i,\pi^a\}   \label{2.1}
\eeq
 dictated by the Gibbs distribution $\exp(-H/T)$ where $H$ is the
Hamiltonian. Since we consider the classical theory, $H$ is just the
classical Hamiltonian for the gauge-Higgs system
\beq
 H = \int d^3 x \left[ \oh E_i^a E_i^a + \oq F_{ij}^a F_{ij}^a
  +  |\pi|^2 + |D_i \phi|^2 + M^2|\phi|^2
  + \l |\phi|^4 \right]
\label{2.2}
\eeq
and in the temporal gauge it should be supplemented by
Gauss constraint in the form
\beq
D_i^{ab} E_i^b = ig(\phi^\dg \tau^a \pi-\pi^\dg \tau^a \phi ).   \label{2.3}
\eeq
After having chosen a configuration according to Gibbs distribution we let
the system evolve according to the classical equations of motion:
\bea
\frac{dA^a_i}{dt}& = & \frac{\delta H}{\delta E^a_i}
\;\;= E^a_i   \label{2.4} \\
\frac{dE^a_i}{dt} &=& -\frac{\delta H}{\delta A^a_i}.
\label{2.5}
\eea
If the system (i.e. in practise the lattice) is sufficient large
the temperature will be approximately constant in this micro-canonical
simulation.

For details about the discretization of the above classical equations
and choice of (lattice) coupling constants we refer to \cite{aaps}.
Here it is sufficient to say that the simulations were performed on
a $16^3$ lattice and that the bare values of the coupling constants
for the gauge fields and the Higgs fields were choosen such that
the tree-values of the theory corresponded to being in the broken phase
with the Higgs mass equal the $W$-mass and such that one sphaleron fits
onto the lattice. As noticed in \cite{aaps} these constraints on the coupling
constants force us for the given lattice size to work at a temperature
where the symmetry is actually restored.

In the symmetric phase we can now measure the change in Chern-Simons number
\beq
N_{cs}(t)-N_{cs}(0) = \frac{1}{32\pi^2} \int_0^t dt \int d^3 x
\; F_{\m,\n}^a \tilde{F}_{\m,\n}^a   \label{2.6}
\eeq
The results of a typical measurement is shown in fig.1a.
One identifies  two  ``sphaleron-like'' transitions, of which the last one
seems to correspond to a jump of Chern-Simons number of two units.
Superimposed on these we see a band of short wavelength thermal fluctuations
which carry the main part of the energy.
According to our arguments above the essential features of the sphaleron
transitions should remain unchanged if we strip off the short wave length
thermal fluctuations. We have done this for each of the time-sequence
of configurations we get by solving the classical equations of motion by
iterating for a given configuration the simplest relaxation equation:
\beq
{ \partial\phi \over \partial t} = - {\delta H \over \delta \phi} \; ,\;
{\partial A \over \partial t}=-{\delta H \over \delta  A} \label{2.7}
\eeq
This technique is well known from the study of lattice instantons and
monopoles \cite{bt,ilmss,ls,hands}.   The results for the configurations
of fig.1a  are shown in fig.1b. Each configuration used in fig. 1a
has been subjected to six cooling sweeps of the kind given in \rf{2.7}.
By this process the energy stored in the gauge fields drops by almost a
factor 80, in agreement with previously obtained results for instantons
and monopoles. The picture of  sphaleron transitions is considerable
sharpened and the fact that the transition survives this cooling
shows at least that the assumption of an effective decoupling of
the short and long wave length thermal fluctuations when we discuss
sphaleron-like transitions is internally
consistent\footnote{It does of course not prove that that we can actually
replace the short wavelength {\it quantum} thermal fluctuations by
short wavelength {\it classical} thermal fluctuations.}. The Higgs field
relaxes much slower and there seems to be a large degree of decoupling
between the Higgs field and the gauge field in the symmetric phase. In fact
we get essentially the same Chern-Simons picture as in fig. 1b if we ignore
the Higgs field and only relax the gauge field. Insensitivity with respect
to the Higgs field has also been observed in the study of
monopole-like configurations \cite{hands}.

\newsection{Spectral flow}

Our aim is to provide further evidence that the sphaleron-like transitions
have the essential characteristica of the continuum, relevant for the
anomaly. One important feature, contained in the Atiyah-Patodi-Singer
index theorem \cite{aps} and explained in detail for instance by
Christ \cite{christ} is that the spectral flow of eigenvalues of the
time dependent Dirac operator $H_D(A(t)$ is directly related to
the change in Chern-Simons number if we consider a continuous time sequence
of $SU(2)$ gauge potentials $A(t)$ starting in one gauge vacuum at
$t= -\infty$ and ending up in a neighbour one at $t=+\infty$. In fact, if
we consider fermions of a given chirality the change in Chern-Simons number
is equal the number of times eigenvalues of $H_D(t)$ cross zero from
below minus the number of times  eigenvalues cross zero from above (the
value zero has no special status in this context, but is conveniently chosen
since the vacuum for $A_i=0$ is identified with the filled Dirac sea).

One interesting point in the present situation is that we have at
no time a true interpolation from one gauge {\it vacuum} to a neighbour one,
since we are working at a finite temperature.
We expect that the Atiyah-Patodi-Singer
index theorem may be used in this more general context, just moving from
one  gauge field configuration to a neighbour one connected by a large gauge
transformation, but have not attempted to provide a rigorous proof of this
conjecture. In practise we try, as already mentioned, to use as smooth
(i.e. cold) configurations as possible.

In the following we will be satisfied with verifying that the crossing
of eigenvalues at zero  is closely linked to the change of Chern-Simons number
as measured in the most ``naive'' way, as described above. Since the
concept of chirality on the lattice is non-trivial \cite{gs,hands}
we will not at the present stage try to unravel the exact counting and
assignment of chirality of the individual modes.

We now turn to the measurement of eigenvalues of the Dirac operator on the
lattice. We have found it most convenient to use the staggered fermion
formalism. The three-dimensional staggered fermion Dirac equation reads:
\beq
\sum_{i=1}^3 \oh i \eta_i (n)\left[ U_i \chi(n+\hat{i}) -
U^\dagger_i (n- \hat{i}) \chi(n-\hat{i}) \right] = E \chi(n) \label{3.1}
\eeq
where $U_i(n)$ are the $SU(2)$ gauge field variables living on the
links $i$ located at lattice points $n$. The formal relation to  a
continuum gauge connection is $U_i(n)= \exp (iA_i(n))$. The fermion field
$\chi(n)$ is a one-component spinor  and an $SU(2)$ doublet.
$\eta(n)$ is the Kawamoto-Smit phase
$(-1)^{n_1+ \cdots +n_{i-1}}$. We use antisymmetric boundary conditions,
in which case there are no zero modes in the free field case and
it is  easier to identify an eigenvalue crossing zero.

In fig. 1c we have shown the time evolution of the lowest {\it positive}
eigenvalue for the
gauge field configurations which were responsible
for the change in Chern-Simons
number shown in fig. 1b. These are the configurations  which are
cooled and thereby have lost most of the short range thermal fluctuations.
However, even in the case where the full thermal fluctuations are present
we  see essentially the same picture.  We have chosen to show in fig. 1c
the eigenvalue from fig. 1b only because it is easier to identify
the Chern-Simons number on this figure.

It is seen that the first diving of an eigenvalue to zero coincides with the
first sphaleron transition of fig. 1b. We have not shown the higher
eigenvalues of the Dirac equation, but they do not  get below 0.15. They show
however a similar (relative) diving as the lowest mode, and they should,
since the choice of zero as the point to count the crossing of eigenvalues
is arbitrary, as mentioned above.

The situation is more complicated for the next change of Chern-Simon number
seen in fig. 1b since it consists of two successive jumps. Accordingly
we see indeed in fig. 1c the diving of two eigenvalues. A closer look
at fig. 1b reveals that the Chern-Simons number
 seems to stop for some time a the value 1/2, precisely the value
 of the sphaleron, which we know has a zero mode \cite{callias,ringwald}.
Notice that this plateau  at 1/2 is present both before and after cooling and
from this point of view should be taken seriously as a continuum
configuration.
Its presence is very clearly reflected in the behaviour of the lowest
eigenvalue. As long as the configuration stays at a value of $N_{cs}(t)
\approx 1/2$ the eigenvalue is close to zero as we have shown
in fig. 2. An obvious interpretation
of this behaviour could be that the system spends   some  time in a
sphaleron like configuration without being able to make up
its mind into which valley (gauge vacuum) to fall. If that is the
case the word ``sphaleron'' is indeed  appropriately chosen for this
configuration.

\newsection{Energy lumps}

In order to understand better the nature of the gauge field configurations
responsible for the change in the Chern-Simons numbers and the spectral
flow of eigenvalues in fig. 1
we have recorded the actual energy distribution of the gauge fields.
It was impossible to see any clear picture when all thermal fluctuations
were included. Only for the cooled configurations did a clear picture
emerge. (Again this is consistent with the experience from instantons
and monopoles).
The average energy of the gauge fields after cooling is around
0.039 and we have chosen to show the energy concentrations of the
gauge field at two times: before the first sphaleron-like  transition and
in the middle of the transition, where the eigenvalue of the Dirac operator
is close to zero (0.002). In both cases we show a sequence of 3D pictures
where regions of space occupied by cubes have an energy density above a certain
threshold (fig. 3). The three thresholds chosen are 0.06, 0.07 and 0.08.
Fig. 3a-3c illustrate a typical situation before the rapid change in
Chern-Simons number while fig. 3d-3f show the spatial distribution
in the middle of the transition.  We see a marked difference
in the concentration of gauge-field energy for the two situations. In the case
of the zero eigenvalue one can talk about a genuine extended object while
the other situation reflects typical fluctuations in the energy density
which will always be present if we pick a random configuration. We see that
the sphaleron-like configurations in no way can be considered as nice
symmetric configuration, but this is not really to be expected.
As a dual representation we show in fig. 4 the energy density
for the extended  object present in the case where the eigenvalue
is zero (fig. 3d-3f) in a plane where the concentration in the core of
the energy lump is high.

\newsection{Discussion}

We have shown that the configurations responsible for the
change in Chern-Simons number  on the lattice indeed seem to share the
characteristica of true continuum configurations  with the same properties.
They have considerable spatial extension and eigenvalues of the Dirac
equation will dive to zero a some point during the change of the Chern-Simons
number. For the sake of clarity we have in this paper
concentrated on a particular
simulation and two sphaleron like transitions, but we have performed a whole
sequence of such simulations and in the process of analyzing the results
we have in fact seen an even more detailed relationship between
the three quantities: the Chern-Simons number, the energy lumps and the
eigenvalues. One type of behaviour is the following:
when performing some  cooling steps we still find some
fluctuations in Chern-Simons numbers which can not be classified as
jumps of order one and where two eigenvalues will dive towards zero, somewhat
displaced. A possible interpretation is that we have two sphaleron like
configurations, the appearance slightly displaced in time, with values of
$N_{cs}(t)$ of opposite sign. Another type of behaviour which we have
observed is one where an eigenvalue seemingly crosses zero and comes back again
or which moves close to zero and then return. At the same time the Chern-Simons
number will move close to approximately 1/2 and return to zero again.
It is again tempting to view this as a gauge field configuration which moves
to a sphaleron-like configuration and then return back to  the same
vacuum-like configuration from which it originated.  In this way it seems
possible to map out in detail the movement of gauge field configuration
relevant for the change of Chern-Simons number and to get a detailed
knowledge of the mechanism by which thermal fluctuations are able to
co-operate and create sphaleron-like configurations.

Many things could be improved: One could obviously gain a lot if it was
possible to use significantly larger lattices\footnote{Unfortunately
we do not have the computer facilities needed for such calculations.}.
It would then be possible to
go to smaller temperatures and smoother configurations, and maybe
even to address the same questions in the region
where the symmetry is broken, but where the
classical transitions from one vacuum to another are not yet suppressed.

Another important improvement  if one wants to
dig into a more detailed investigation of the flow of eigenvalues
is the chirality of the eigenmodes. A suitable definition has been given
in \cite{gs,hands}.

A final very important question which remains to be addressed is the role of
the
Higgs field. This role is still obscure to us.
As already reported long ago \cite{als}
it seems that the Higgs field in the symmetric phase partly decouples from the
gauge field. Since we effectively are in the symmetric phase the Higgs field
will fluctuate and have zeroes, and when we iterate a few times the relaxation
equations it still seems somewhat
decoupled from the gauge field.
At smaller temperatures one would expect a
strong coupling of the phases of the Higgs field and the
gauge field. Since the real fermions in the electroweak theory couple
to the Higgs field it is indeed important to understand the role of this
field. In the continuum the situation is only clear in the broken
phase. If a field configuration in the broken phase
(gauge fields and Higgs fields) interpolates in time between two
non-equivalent vacua the Higgs field {\it has} ( by purely topological reasons)
 to develop a
zero in between the vacuum configurations where the magnitude of the
Higgs field is constant. This zero of the Higgs field is precisely what
is present in the sphaleron configuration and it is this zero which allows
the Dirac operator, coupled both to gauge- and Higgs fields of the sphaleron,
to have a normalizable energy eigenmode
with eigenvalue zero, representing precisely the crossing of zero of the
energy levels\cite{callias,ringwald}. We do not know of any detailed
investigation of the
role of the Higgs field in the symmetric phase.

\vspace{12pt}

{\bf Acknowledgement} It is a pleasure to thank Morten Laursen for
discussions and him and J. Vink for providing us with a program for
the calculation of  eigenvalues of staggered fermion Dirac equation.

\vspace{12pt}

\addtolength{\baselineskip}{-0.20\baselineskip}

\newpage

\begin{center}

{\large \bf Figure Caption}

\end{center}

\vspace{24pt}

\begin{itemize}
\item[Fig.1]
Measurements of the Chern-Simons as a function of time. Fig. 1a and
shows time  evolution with thermal fluctuations included, while
fig. 1b shows the same time evolution with the thermal
fluctuations partly stripped off. Fig. 1c shows the lowest eigenvalue
for the configurations which are responsible for the Chern-Simons
numbers shown in fig. 1b.

\item[Fig.2]
Fig. 2a shows in more detail the behaviour of the Chern-Simons number
in the neighbourhood of the second sphaleron transition recorded in fig. 1b.
Fig. 2b shows details of the lowest eigenvalue corresponding to fig. 2a.

\item[Fig.3]
Fig. 3a-3f are
3D pictures of the regions in space where the energy concentration
is larger than 0.06, 0.07 and 0.08  for two configurations
before and in the middle of the first sphaleron transition
shown in fig. 1b.
The first three figures refer to a configuration before the transition
and for which there are no small eigenvalues, while the last three figures
refer  to  a configuration in the middle of the transition. This configuration
has an eigenvalue of the Dirac operator which is approximately zero.

\item[Fig.4]
An energy level plot in the $x-y$-plan of fig. 1c at $z$-height where the
energy concentration in the core is high.

\end{itemize}

\end{document}